\documentstyle[aps,epsf]{revtex}                         % submission          %
\draft
\begin{document}
\twocolumn[\hsize\textwidth\columnwidth\hsize
     \csname @twocolumnfalse\endcsname

%\preprint{}%

%------------------------------------------------------------------------------%
%       T I T L E                                                              %
%------------------------------------------------------------------------------%

\title{Understanding Spin: 
       the field theory of magnetic interactions.
       }

\author{W. A. ~Hofer}
\address{
         Dept. of Physics and Astronomy, University College London \\
         Gower Street, London WC1E 6BT, E-mail: w.hofer@ucl.ac.uk}

\maketitle

\begin{abstract}
Spin is commonly thought to reflect the true quantum nature of microphysics.
We show that spin is related to intrinsic and field-like properties of 
single particles. These properties change continuously in external magnetic
fields. Interactions of massive particles with homogeneous and inhomogeneous 
fields result in two discrete particle states, symmetric to the original one.
We analyze the difficulties in quantum mechanics to give a precise spacetime 
account of the experiments and find that they arise from unsuitable analogies 
for spin. In particular from the analogy of an angular momentum. Several 
experiments are suggested to check the model against the standard model in 
quantum mechanics.
\end{abstract}


\vskip2pc]

\section{Analogies and understanding}

It has frequently been stressed that the mathematical form of quantum physics 
can be deduced from group theoretical considerations \cite{mirman76,fano96}. 
A case was then made, that the linearity of quantum mechanics (= validity of
{\it superposition}), together with the necessary symmetry operations,
provides a full understanding of the theory. Even though this is true 
for a mathematician, it is not necessarly true for a physicist. The work of a 
mathematician is done, when the work of a physicist only begins. Let us 
call this part of theoretical physics, which differs from pure mathematical 
reasoning, {\em understanding} the physical theory. Understanding, in this 
sense, means a clear conceptual picture, why an entity must be described the way 
it is described, why these descriptions operate, the way they operate, and 
why they lead to the results they provide. 

One of the most useful logical figures in understanding physics is 
the {\it analogy}. A difficult phenomenon is, via analogy, referred to
a less difficult and more easily understandable one. The drawback
of analogies is, they are not determined by logical necessity. Note that
we use this term rather than the mathematically more precise {\it mapping}
because of its connotations. While mapping only means that a series of
experimental results can be mapped onto a theoretical model, a suitable
analogy equals a consistent description of the physical processes involved.
The two terms differ in their epistemological implications: whereas
analogy is related to realism, mapping implies an essentially positivist
view. For this reason we prefer analogy. We shall argue, in this paper, 
that the usual analogies for {\it spin} are wrong. Spin should be understood
as something, the evaluation procedure in quantum mechanics does to the 
mathematical representation of a system. Incidentally a point, Bohr
would have subscribed \cite{bohr35}. But we shall also describe the physical 
qualities of {\it single} particles, acted upon by external magnetic fields.
And give a precise spacetime picture of the evolution of a particle's 
magnetic properties. A description, commonly denied in standard frameworks.

The analogy for spin is based, like most analogies in quantum mechanics, 
on classical mechanics and electrodynamics. In particular, as we will 
prove, on a misleading combination of both to understand magnetic 
properties of matter. It is derived from the {\it magnetic moment} 
of current distributions in electrodynamics \cite{jackson99}. 
A current carrying loop, placed
into a magnetic field, experiences a force along the field gradient.
This force is proportional to the magnetic moment of the current loop.
The implicit concept below the mathematical derivation of this force
is the concept of Ampere's molecular currents. In case of single
particles the current carrying loop is replaced by the mechanical 
picture of a point charge in orbit around a defined center. This is
the reason that in the original papers spin was usually called the
''spin angular momentum''.

With this analogy in mind Stern and Gerlach \cite{stern24}
performed experiments on the
magnetic moment of silver atoms. The results did not agree with the simple
classical picture. So they concluded that the simple classical
picture, based on the described analogy, must be wrong. The answer
to this problem, in
the view of quantum mechanics, was given by Goudsmit and Uhlenbeck
\cite{goudsmit25a,goudsmit25b}: the 
electron possesses an intrinsic angular momentum, which reveals itself in
these measurements via its magnetic moment. This answer was generalized
into an impressive mathematical formalism, not least by Pauli 
\cite{pauli40}. It is
still the main answer, physicists accept as ultimately true. The problem,
though, which remained, is the understanding of this intrinsic angular
momentum. All attempts to refer it to some motion of 
a structured particle with a point-like charge did not convincingly
remove the objection that this rotation seemingly has no direction in
space. Then how can it be a rotation? (for a review of electron models
trying to establish spin on this basis see \cite{keller97}). 

At this point we have two possibilities: (i) Either leave it at that 
and accept the numerical recipes as useful. (ii) Try to create a 
different, more suitable analogy. Which is, trying to understand spin.
Even though the first approach has been employed in the past, the 
author does not consider it, ultimately, a physical approach: it 
leaves the problem of interpretation aside and thus is no more 
than mathematics. It is not enough for physics. 

Quite generally, the approach
to physics in the last, the 20th century, to leave all problems of 
interpretation aside and focus on mathematics alone, is inefficient.
Because then only one route remains to new insights: mathematical
intuition. Given that this is the most formal and hence least 
versatile intuition in terms of human imagination, the procedure 
in itself is questionable. Considering, in addition, that a 
precise image of a process can cover, within instants, developments 
it takes major computational efforts to reproduce numerically, the 
whole idea of a natural science based only on mathematical structures 
seems utterly absurd. All considered, it is not quite as absurd as
that, because mathematical consistency has the 
advantage of greater generality. But there is a limit, as Aristotle
remarked some two thousand years ago: the most general concepts 
are also the emptiest. And the ultimate structure of mathematical
physics, the most general theory, will be devoid of any 
meaning. All it will cover is the structure of mathematics itself.
This is quite compatible with the basic assumption in what became 
known as the {\it Copenhagen Interpretation} \cite{bohr59}: 
there is no independent reality. 

In this paper we will base a theory of magnetic interactions
on just such an independent reality, the intrinsic and field-like 
properties of single particles. In section \ref{sec_micro} we shall
briefly review the theoretical foundations of the treatment. Section
\ref{sec_inter} treats the interaction of neutral particles (neutrons,
but also atoms) with homogeneous and inhomogeneous magnetic fields. 
Then, in section \ref{sec_measure} we shall briefly discuss measurements
and statistics, while section \ref{sec_spin} establishes an actual 
meaning of the spin concept in quantum mechanics. 

\section{Foundations of Microdynamics}\label{sec_micro}

The theoretical framework in this paper is based on a 
{\it strongly objective} \cite{despagnat89} theory of microphysics.
Strong objectivity means that the entities we deal with in microphysics
are assumed to have a reality and properties, which are completely
independent of any measurement. These entities exist separate from any description, 
and even though their qualities change in interactions (that is, what we 
call a {\it measurement}), the qualities as such prevail also without
measurements. In this sense the theory gives a definite answer to Mermin's
question \cite{mermin85}: '' Is the moon there when nobody looks?'',
and the answer is: Yes, certainly.

\begin{figure}
\begin{center}
\epsfxsize=1.0\hsize
\epsfbox{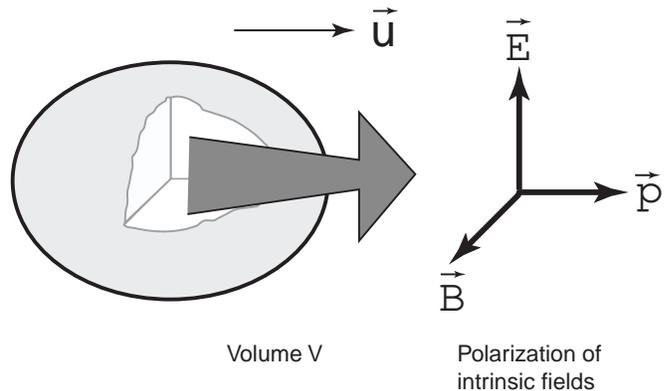}
\end{center}
\vspace{0.5 cm}
\caption{
         The intrinsic fields of a particle in motion. The particle
         is extended over a volume $V$, which is not specified. The
         longitudinal velocity ${\bf u}$ and momentum density ${\bf p}$
         are perpendicular to the transversal intrinsic fields 
         ${\bf E}$ and ${\bf B}$.         
        }
\label{fig001}
\end{figure}

The theory has its foundations in the intrinsic properties of moving particles. 
It was called, for obvious reasons, the Theory of Microdynamics (MD) 
\cite{micro_web}.
Magnetic interactions and the analysis of spin will be treated from the
viewpoint of the MD model of particles \cite{hofer98}. The main 
features of this model are (see Fig. \ref{fig001}):

\begin{itemize}
\item
Particles are extended structures, the actual extension (= the volume of the
particle) is usually irrelevant. This feature is due to the mathematical
structure based on local differential equations.
\item
Particles are described by their intrinsic features alone: the
variables of the description are mass density $\rho ({\bf r},t)$,
frequency $\nu$, the longitudinal momentum density ${\bf p}({\bf r},t)$,
and the transversal intrinsic fields ${\bf E}({\bf r},t)$ and 
${\bf B}({\bf r},t)$.
\item
The development of these intrinsic properties is governed 
by linear field equations connecting the variables.
\item
Scalar fields and dynamics of particle propagation are related
via the concept of {\it dynamic charge} \cite{hofer00a}. This
concept is similar in spirit to the harmonic oscillator in 
quantum mechanics, but it remains firmly within a continuous
and field theoretical framework.
\end{itemize}

The intrinsic fields and their properties are not statistical 
features of single particles, but essentially deterministic. The
statistical ensembles, which also in MD arise in experiments, have
two origins \cite{hofer00c}:

\begin{itemize}

\item
They are an expression of limited control of the experimental
conditions. In the same sense as e.g. in the de Broglie-Bohm theory
\cite{holland93}, the phase of intrinsic fields of a single particle
is in general unknown.

\item
The ensembles are also an expression of limited knowledge about the 
interaction. From the viewpoint of MD also electrodynamics is a
statistical theory \cite{hofer98}. This means that fields
in electrodynamics have components interacting with a particle's
field-like properties, and components, interacting with its
kinetic properties. A way to describe these (separate) components
is using real and imaginary fields \cite{hofer00b}.

\end{itemize}

The equations necessary for the presentation will be stated in 
later sections. It should be noted that standard quantum
mechanics (the Schr\"odinger equation) as well as standard 
electrodynamics (the Maxwell equations) can be referred to 
the field equations of intrinsic particle properties \cite{hofer98}. 
Therefore the treatment is general in terms of both theories. As will
be seen presently, it allows an analysis of interaction processes
from a field theoretical point of view and still recovers the
essential results of quantum mechanics. In this sense the
whole concept is well beyond a purely statistical interpretation
of quantum mechanics and in fact a hidden variable theory.
The hidden variables, in this theory, are the intrinsic 
fields and densities.

\section{Interaction of fermions with magnetic fields}\label{sec_inter}
\subsection{Homogeneous fields}

\begin{figure}
\begin{center}
\epsfxsize=1.0\hsize
\epsfbox{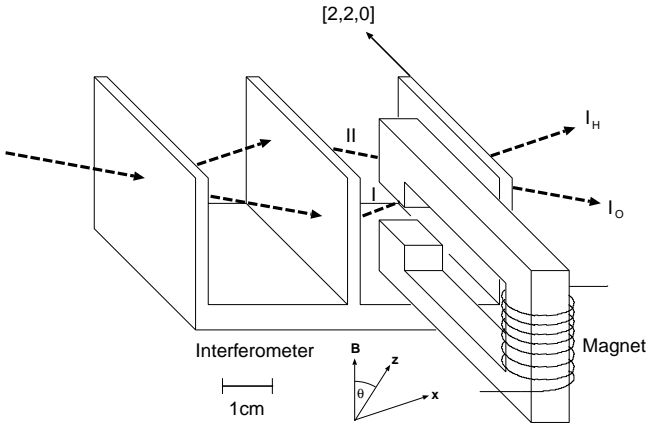}
\end{center}
\vspace{0.5 cm}
\caption{
         Neutron interferometer according to Rauch [20].
         The coherent Neutron beam is split at the first lobe and 
         recombined after the third. A magnetic field is applied
         to one partial beam (I). The effect is a shift of the 
         interference pattern after recombination.
         }
\label{fig002}
\end{figure}

We consider the motion of a beam of neutral particles in a homogeneous 
magnetic field. The trajectory of the particles is not affected.
But recombining this beam with a reference beam not subjected to the
magnetic field shows a changed interference pattern 
\cite{zeilinger81,rauch92} (see Fig. \ref{fig002}). 
The question is: why?

The particles are neutral, therefore an interaction between the 
external magnetic field and particle charge is excluded. The
change of properties then must be due to intrinsic features
not readily observed in standard experiments. In the conventional 
model this intrinsic feature is the spin of a particle, e.g.
a neutron (half-spin) or photon (integer spin). 
In most textbooks spin is treated as a somewhat abstract 
variant of angular momentum (see e.g. \cite{merzbacher98}).
This concept can be traced back to the original papers by 
Uhlenbeck and Goudsmit, where the authors concluded that 
\cite{goudsmit25a}: 

{\it Das Elektron rotiert um seine eigene Achse
mit dem Drehimpuls $\hbar/2$. F\"ur diesen Wert des Drehimpulses
gibt es nur 2 Orientierungen f\"ur den Drehimpulsvektor.}

The obvious objection, that in macroscopic systems no angular momentum 
is known with just two orientations in space has been brushed aside 
with the remark, that this is due to the ''quantum nature'' of spin.
What the difference between this ''quantum nature'' of the ''spin
angular momentum'' and the realistic nature of a macrophysical 
angular momentum actually is, has never been
conclusively explained. And, as recent research revealed (see the
introduction), it cannot be explained in a manner consistent
with experimental facts. In this context it should be noted that
the explanation of spin 1/2 by the Dirac equation is not correct. 
Because, as Gottfried and Weisskopf pointed out \cite{macgregor92}:

{\it At one time it (the Dirac equation) was thought to 'explain'
the spin s = 1/2 of the electron, but we now know that this is not
so. Equations of the Dirac type can be constructed for any s. At 
this time we have no understanding of the remarkable fact that 
the fundamental fermions of particle physics ... all have spin 1/2.}

The model of extended particles does not provide a model of spin in
its usual sense, all that exists are transversal intrinsic
fields, called {\it electromagnetic} fields for convenience. The fields
obey the Maxwell equations, the relevant equations are the following
\cite{hofer98}:

\begin{eqnarray}
\frac{1}{u^2} \frac{\partial {\bf E} ({\bf r},t)}{\partial t} = 
\nabla \times {\bf B}({\bf r},t)  \\
\qquad - \frac{\partial {\bf B} ({\bf r},t)}{\partial t} = 
\nabla \times {\bf E} ({\bf r},t) 
\\ \phi_{EM}({\bf r},t) = \frac{1}{2} \left( 
\frac{1}{u^2} {\bf E}^2 ({\bf r},t) + {\bf B}^2 ({\bf r},t)
\right) \end{eqnarray}

Here $u$ is the velocity of the particle and $\phi_{EM}$ the energy
density of the electromagnetic field. We assume that the particle moves 
along $x$, the electromagnetic fields shall initially be polarized along
the $y$ and $z$ axis of our coordinate system.

\begin{eqnarray}
{\bf E} ({\bf r},t) = (0, E_{0}, 0) e^{i(k_{0}x - \omega_{0}t)} \nonumber \\
{\bf B} ({\bf r},t) = (0, 0, B_{0}) e^{i(k_{0}x - \omega_{0}t)}
\end{eqnarray}

We consider the change of intrinsic fields due to an external magnetic field
$B_{ext}$. The dynamic changes of the field environment, when the particle
approaches the magnet are simulated by a finite interval $\tau$ taken for the
field to be fully switched on. The field vector of the external field shall
be:

\begin{equation}
{\bf B}_{ext} = (0, - \sin \vartheta, \cos \vartheta) B_{ext}
\end{equation}

For a linear increase of the field during $t \in [0, \tau]$ the internal
fields at $\tau$ will be:

\begin{eqnarray}
E'_{y} & = & E_{0} e^{i(k_{0}x - \omega_{0}t)} - B_{ext} \cos \vartheta 
\frac{x}{\tau} \nonumber \\
E'_{z} & = & - B_{ext} \sin \vartheta \frac{x}{\tau} \\
B'_{y} & = & - B_{ext} \sin \vartheta \nonumber \\
B'_{z} & = & B_{0} e^{i(k_{0}x - \omega_{0}t)} + B_{ext} \cos \vartheta 
\nonumber
\end{eqnarray}

\noindent
The amplitudes of the electromagnetic fields are related by \cite{hofer98}:

\begin{equation}
E_{0} = u_{0} B_{0}
\end{equation}

Then the change of the electromagnetic energy density due to the external 
magnetic field is given by:

\begin{equation} \label{1006}
\delta \, \phi_{EM} (\tau, x) = \frac{1}{2} \left(
\frac{B_{ext}^2}{u_{0}^2}  \, \frac{x^2}{\tau^2} + B_{ext}^2 \right)
\end{equation}

One key feature of this new model of particles is the relation
between mass oscillations (or changes of the density of mass) and 
scalar fields. The relation is described 
by the differential equation \cite{hofer00a}:

\begin{equation}
\triangle \phi ({\bf r},t) + \frac{\partial^2 \rho ({\bf r},t)}{\partial t^2} = 0
\end{equation}

In this one-dimensional case the integration of the equation is 
straightforward and the result obtained for the changed density
of mass $\delta \rho$ and kinetic energy density 
$\delta \phi_{K} = \delta \rho u^2$ amounts to:

\begin{eqnarray}
\delta \rho (\tau) & = & - \frac{1}{2} \frac{B_{ext}^2}{u_{0}^2}  \\
\delta \phi_{K} (\tau) & = & \delta \rho (\tau) 
\cdot u_{0}^2 = - \frac{1}{2} B_{ext}^2
\end{eqnarray}

In general electromagnetic fields must be described by complex vectors.
This feature derives from the simultaneous existence of field components 
and kinetic components of particle energy. A thorough account of it
will be given at the end of this section. Denoting the real and the 
imaginary part of the magnetic field by:

\begin{equation}
B^{+} = \Re (B_{ext}) \quad B^{-} = \Im (B_{ext}) \quad
|B^{+}| = |B^{-}| = B_{E} 
\end{equation}

\noindent
we get for the change of the kinetic energy density in a magnetic field 
the two symmetric solutions:

\begin{equation}
\delta \phi_{K}^{\pm} (\tau) = \mp \, \frac{1}{2} B_{E}^2
\end{equation}

\begin{figure}
\begin{center}
\epsfxsize=1.0\hsize
\epsfbox{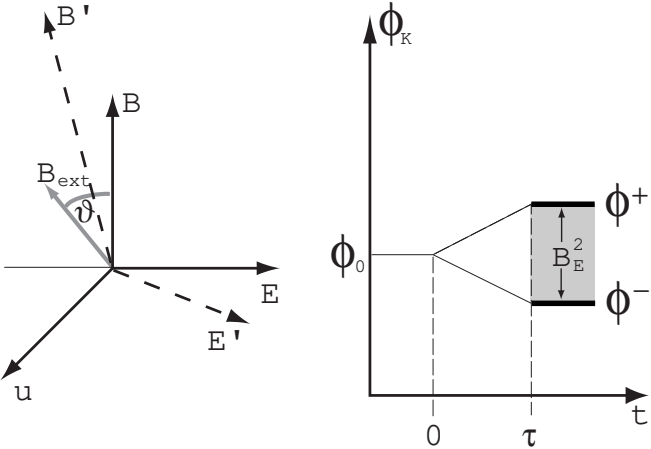}
\end{center}
\vspace{0.5 cm}
\caption{
         Intrinsic fields of the particle and external magnetic
         field. The external field $B_{ext}$ is perpendicular
         to the direction of motion and in an angle $\vartheta$ with
         the particle's intrinsic field $B$ (left). 
         The kinetic energy density is changed to two
         separate levels $\phi^{+}$ and $\phi^{-}$. The difference
         between these two levels is equal to $B_{E}^2$. Note that the
         change of the energy occurs continuously within the 
         interval $t \in [0,\tau]$ (right).   
         }
\label{fig003}
\end{figure}

The intrinsic fields of the particle and the changes due to the interaction
with the magnetic field are shown in Fig. \ref{fig003}. 
This change affects the two contributions to
the energy, the field components and the kinetic components, in 
opposite ways. The total energy of the particle remains unchanged. This is,
incidentally, also a requirement of electrodynamics. The requirement is 
not fulfilled in quantum mechanics. A fact, which so far seems to have
escaped wider notice. Because strictly speaking it contradicts the field
equations of electromagnetic fields. 

Quite generally, the energy of a particle in a magnetic field is conserved.
This is a consequence of the Lorentz force equation:

\begin{eqnarray}
{\bf F}_{L} = q {\bf v} \times {\bf B} \qquad
dW = {\bf F}_{L} d {\bf r} \nonumber \\
{\bf F}_{L} \perp \frac{d {\bf r}}{d t} \Rightarrow dW = 0
\end{eqnarray}
 
If energy is transferred at all to the particle, then the medium of
transfer must be the magnetic field. From the energy integral:

\begin{equation}
W_{B} \propto  \int {\bf B}^2 \, d^3 r 
\end{equation}

it follows that this can only be accomplished by a change of the field itself:

\begin{equation}
\triangle W_{B} \propto \int_{R} (\triangle {\bf B})^2 \, d^3 r
\end{equation}

where $R$ is the region of interaction. This requires that the absolute 
value and not only the the direction of the field changes due to the 
interaction. Which is only possible, if there is a drain of the field in
the same region. And this, in turn, contradicts the source equations of
magnetic fields:

\begin{equation}
\nabla {\bf B} = 0
\end{equation}

Therefore any transfer of energy from the field to the particle is 
not possible within the framework of electrodynamics. In other words,
the QM concept of spin violates the energy principle in all cases,
where it predicts a change of particle energy due to magnetic fields.

In the current treatment the magnetic field only changes the distribution 
between the two energy components. The changes of the energy distribution
of the particle result from the dynamic changes of the field environment. 
In this sense a constant field does no longer affect the particle's motion.
This constant field is equal to the second term in equation (\ref{1006}).
Taking consequently only the first term of Eq. (\ref{1006}) we get for the 
components and the total change of energy density of the particle:

\begin{eqnarray}
\delta \phi_{EM}^{\pm} (\tau) & = & \pm \frac{1}{2} B_{E}^2 \nonumber \\
\delta \phi_{K}^{\pm}  (\tau) & = & \mp \frac{1}{2} B_{E}^2 \\
\delta \phi_{T} (\tau) & = & \delta \phi_{EM}(\tau) +
\delta \phi_{K} (\tau) = 0 \nonumber
\end{eqnarray}

The change of the particle's density of mass does not leave the dynamic
properties of motion unaffected. In particular the phase velocity in the
magnetic field will be changed. This can be shown using the continuity
equation:

\begin{equation}
\frac{\partial \rho ({\bf r},t)}{\partial t} + 
\nabla {\bf p} ({\bf r},t) = 0 
\end{equation}

At an arbitrary point (x,t) along the change of the external field the 
changes of $\rho$, density of mass, and $u$, phase velocity of the particle's
wave (which is equal, in this model, to the mechanical velocity), are 
described by:

\begin{eqnarray}
\rho (t) & = & \rho_{0} + \delta \rho (t) =
\rho_{0} - \frac{1}{2} \frac{B_{ext}^2}{u_{0}^2} \cdot \frac{t^2}{\tau^2}
\nonumber \\
u (x) & = & u_{0} + \delta u (x)
\end{eqnarray}

\noindent
Omitting terms of second order the changes are consequently:

\begin{equation}
\frac{\partial \delta \rho (t)}{\partial t} = - \rho_{0}
\frac{\partial \delta u (x)}{\partial x}
\end{equation}

\noindent
Integrating we get for the change of the phase velocity at an arbitrary
point (x,t):

\begin{equation}
\rho_{0} \delta u (x, t) = - \int_{0}^{x} dx' 
\frac{\partial \delta \rho (t)}{\partial t} = \frac{1}{2}
\frac{B_{ext}^2}{u_{0}^2} \frac{x t}{\tau^2}
\end{equation}

\noindent
And at the moment $t = \tau$, with $x = x_{1}$ and $x_{1}/\tau \approx u_{0}$
this amounts to:

\begin{equation}
u_{0} \delta u (\tau) \approx 
\frac{1}{2} \delta \, u^2(\tau) = \frac{1}{2} \, \frac{B_{ext}^2}{\rho_{0}}
\end{equation}

Again, generalizing the expression for complex fields we obtain two symmetric
solutions for the phase velocity in the magnetic field. The velocity 
difference, and thus the phase shift in the field is linear with the
amplitude $B_{E}$ \cite{rauch92}:

\begin{equation} \label{1021}
\delta u (\tau) = \pm \frac{B_{E}}{\sqrt{\rho_{0}}}
\end{equation}

The change of the phase velocity leads to a phase shift of the beam component
in the magnetic field compared to the phase of the reference beam. This shift
is linear with the amplitude of the magnetic field. 
Summing up the results in homogeneous fields we can say that the treatment
allows to recover all features of the result in quantum mechanics
(two discrete and symmetric solutions for the kinetic energy and the
phase velocity, phase shifts linear with the amplitude of the field), 
while it gives a fully deterministic and local treatment 
of the process. An application of the result to atomic physics, which we 
will give in a follow up publication, accounts for the main features of the 
{\it normal Zeeman effect} in atomic physics.

\subsection{Inhomogeneous fields: Stern-Gerlach experiments}

In inhomogeneous fields the magnetic intensity and amplitude depends on
the location ${\bf B}_{ext} = {\bf B}_{ext}({\bf r})$. In particular,
in the setup of Stern-Gerlach experiments \cite{stern24}, the magnetic
field changes along the $z$-axis of the coordinate system:

\begin{equation}
{\bf B}_{ext}({\bf r}) \equiv {\bf B}_{ext}(z) \qquad
\frac{\partial {\bf B}_{ext}(z)}{\partial z} \ne 0
\end{equation}

We assume, for the following, that the magnetic field is switched on in 
$t \in [0,\tau]$, and that $\tau$ is independent of $z$. Then the kinetic
energy density $\phi_{K}(\tau, z)$ equals:

\begin{equation}
\delta \phi_{K}(\tau,z) = \frac{1}{2} B_{ext}^2(z)
\end{equation}

Taking the gradient of the energy gives the local change of energy
and thus a Newtonian potential and a Newtonian force acting on the 
particle:

\begin{equation}
F_{z} = - \frac{\partial \delta \phi_{K}(z,\tau)}{\partial z} =
- B_{ext}(z) \frac{\partial B_{ext}(z)}{\partial z}
\end{equation}

Accounting for real and imaginary components of the external field 
$B_{ext}$ gives, as in the homogeneous field, two symmetric solutions.
These solutions describe the force perpendicular to the particle's
path of motion:

\begin{equation}
F_{z} = \pm B_{E}(z) \frac{\partial B_{E}}{\partial z}
\end{equation}

The trajectory of the particle will be affected by this force. Writing 
it as an acceleration acting on the particle's density of mass $\rho_{0}$,
we get finally:

\begin{equation}
\frac{d u_{z}}{d t} = \pm \frac{B_{E}(z)}{\rho_{0}} 
\frac{\partial B_{E}(z)}{\partial z}
\end{equation}

This reflects the standard result in Stern-Gerlach experiments: two symmetric 
solutions on either side of the zero point and in the direction of the field 
gradient.

\subsection{The natural system of units}

For reasons of consistency one has to use {\it natural units} 
in this calculation. In this system of physical units the interface between 
electrodynamics and mechanics is described by $\hbar$, Planck's constant
\cite{hofer00a}. The fields and energies, in natural units, take the following
form (terms of $4 \pi$ generally omitted in the one-dimensional case):

\begin{eqnarray}
\phi & = & \frac{\hbar}{2} \left(
{\bf E}^2 + c^2 {\bf B}^2 \right)
\end{eqnarray}
\begin{eqnarray}
\left[\phi\right] & = & \left[J m^{-3}\right] \nonumber \\
\left[{\bf E}\right] & = & \left[N m^{-3}\right] \\
\left[{\bf B}\right] & = & \left[N m^{-3} s m^{-1}\right] \nonumber \\
\left[\hbar\right]  & = & 1.054 \times 10^{-34} \left[m^4 N^{-1} \right] \nonumber 
\end{eqnarray}

\noindent
Then the results for homogeneous fields, and including $\hbar$, are the
following:

\begin{eqnarray}
\delta \phi_{K}^{\pm}(\tau) & = & \mp \frac{\hbar}{2} \, 
c^2 B_{E}^2 \nonumber \\
\delta u^{\pm}(\tau) & \propto & \pm \frac{\hbar}{2} 
\, B_{E}  
\end{eqnarray}

\noindent
The two symmetric solutions in the inhomogeneous field are:

\begin{eqnarray} \label{1100}
F_{z}^{\pm} & = & \pm \frac{\hbar}{2} \, \frac{\partial}{\partial z}
\,  c^2 B_{E}^2 \nonumber \\
\frac{d u_{z}(\tau)}{d t} & = & \pm \frac{\hbar}{2} \, 
\frac{\partial}{ \rho_{0} \partial z} \, c^2 B_{E}^2
\end{eqnarray}

The main difference between a charged particle and a neutral one in a 
magnetic field is the change of the motion due to Lorentz forces. A
neutral atom, even though its electrons may experience these forces,
is not affected by Lorentz forces. In first approximation one may 
therefore treat the atom as a neutral particle in magnetic fields.
Considering, that electron mass will experience the same shift of
fields and consequently the same changes of kinetic energy in a
dynamic model of atoms \cite{hofer98b,hofer00a}, the results derived must 
also apply for motion of a neutral atom in an inhomogeneous magnetic 
field. The observation of two symmetric trajectories was, of course, 
one of the first experimental results obtained by Stern and Gerlach
\cite{stern24}.   

\subsection{Magnetic interactions: the general picture}

In general, as outlined in the foundations of microdynamics \cite{hofer98},
one has to account for two components of particle energy: the energy
contained in its electromagnetic fields, and its kinetic energy.
Both components together yield the total energy density of a particle,
which is a constant of motion:

\begin{equation}
\phi_{T} = \rho_{T}u^2 = constant
\end{equation}

The amplitude is commonly irrelevant, since it does not show up in the
differential equations describing the system. But the distribution between
kinetic and field components of energy, depending, for a given moment or
location, on the phase of the wave, must be included in the picture.
A standard way to achieve this, is using imaginary components. 
From the viewpont of electromagnetic fields we may add 
imaginary contributions to the - real valued - 
magnetic fields. From the viewpoint of kinetic energy the field components 
are best described by imaginary components to a particles wavefunction,
where $\psi^2 = \rho$, the particle's density of mass. We therefore
get, in general, for $\phi ({\bf r},t)$ and $\rho ({\bf r},t)$ the 
following relations:

\begin{eqnarray}
\phi ({\bf r},t) & = & \frac{\hbar}{2} \left[c \cdot {\bf B} ({\bf r},t)\right]^2 
\qquad {\bf B} ({\bf r},t) = (B_{r} + i B_{i}) {\bf e}_{B} \nonumber \\
\rho ({\bf r},t) & = & \psi^2 ({\bf r},t) \qquad
\psi ({\bf r},t) = \psi_{R} + i \psi_{i}
\end{eqnarray}

The relation between the scalar fields $\phi$ and $\rho$ (see Eq. (9)) 
is then multivalued, it depends on the distribution between the real and the
imaginary components. Physically speaking, it depends on the exact
moment, when the particle begins interacting with the external field.
Since the intrinsic fields of the particle are periodic, it depends
on the phase of the fields. And as there is, in addition, a continuous
shift of energy from the field components to the kinetic components of
a particle's energy and vice versa, the real and complex components
of the wavefunction are in general not zero. The same is true for
the scalar field. The cases we have outlined so far are only
the degenerate ones with:

\begin{eqnarray}\label{3032}
& (I)&  \qquad B_{r} \ne 0 \quad B_{i} = 0 \qquad 
        \psi_{r} \ne 0 \quad \psi_{i} = 0 \nonumber \\
& (II)&  \qquad B_{r} \ne 0 \quad B_{i} = 0 \qquad 
        \psi_{r} = 0 \quad \psi_{i} \ne 0 \nonumber \\
& (III)&  \qquad B_{r} = 0 \quad B_{i} \ne 0 \qquad 
        \psi_{r} \ne 0 \quad \psi_{i} = 0  \\
& (IV)&  \qquad B_{r} = 0 \quad B_{i} \ne 0 \qquad 
        \psi_{r} = 0 \quad \psi_{i} \ne 0 \nonumber   
\end{eqnarray}

A further degeneracy arises from the equivalence of case I, IV and 
case II, III. For an arbitrary relation between intrinsic particle
components and external field components the real and imaginary
parts of the fields comply with:

\begin{equation} \label{3033}
\frac{\hbar}{2}  \, \triangle \left(B_{r} + i B_{i} \right)^2 +
\frac{\partial^2}{c^2 \partial t^2}  
\left(\psi_{r} + i \psi_{i} \right)^2 = 0
\end{equation}

All components, in this general case, are non-trivial solutions of the
differential equation. Note that this is no longer a linear equation.
The non linearity thus enters the framework of microphysics at the
energy transfer from field components to kinetic components and vice
versa. Or, generally speaking, at the transition from electrodynamics
to quantum mechanics. While both theories, in themselves, are linear,
the general theory, including all interactions, is not. In the long
term we expect this feature of microdynamics to provide the most 
interesting deviations from the purely linear frameworks employed in
standard theory. Writing the relation as a system of two nonlinear
equations for real and imaginary components, we get:

\begin{equation}
\frac{\hbar}{2} \,\triangle \left(B_{r}^2 - B_{i}^2 \right) +
\frac{\partial^2}{c^2 \partial t^2} 
\left(\psi_{r}^2 - \psi_{i}^2 \right) = 0
\end{equation}
\begin{equation}\label{3035}
\frac{\hbar}{2} \, \triangle B_{r} B_{i} +
\frac{\partial^2}{c^2 \partial t^2} \psi_{r} \psi_{i} = 0
\end{equation}

It is now easy to see that the degenerate solutions of (\ref{3032}) 
are equivalent to the trivial solutions of Eq. (\ref{3035}).

Note that the results apply to massive particles. In quantum mechanics
massive particles are fermions. In addition, the theoretical framework
is non-relativistic, so it can be applied with confidence only to the
low energy limit of particle motion. Whether a similar picture is
suitable for bosons, cannot be decided at present. Especially,
since the archetypical bosons, photons, are relativistic and
it has been shown, within the current framework, that interactions in
this range are changed due to spacetime transformations in moving
coordinate frames \cite{hofer00d}.

\section{Measurements and statistics}\label{sec_measure}

An common feature of all results, for homogeneous as well as inhomogeneous
fields, is that only two numerical values are generally decisive for the 
outcome of a measurement: the magnetic field (or its gradiant) and the
Planck constant, or rather $\hbar/2$. Even though material parameters of atoms
will enter an evaluation of (\ref{1100}), the functional form remains the same:
a constant multiplied by a gradient and some material parameter. And the constant
is incidentally equal to the spin of a particle. 

In quantum mechanics it is generally thought that the particle is in a state
''spin-up'' or ''spin-down'' or a superposition of both. A measurement with
a Stern-Gerlach device in an arbitrary direction $z$ leaves the particle in one
of either states, but now with reference to the direction $z$. This is the
notorious collapse of the wavefunction. The statistics on quantum mechanics,
even though their change in a measurement is still not explained consistently, 
arise from the qualities of the particles themselves.

In the present picture the statistics arise from the interaction with the magnetic
field. In this sense it is correct to say that the particle, before the 
measurement, does not have a defined property. But does it have defined 
properties {\it after} the measurement? Only, if we concede that the changes
due to the applied fields are permanent. This would require some kind of
resistivity, which we have not introduced. 

In principle, this feature can be added to the model by some kind of internal
friction. However, at this point we thought it sufficient to introduce the
theory without it, because it is not clear, whether the feature is at all 
needed. Only experiments are suitable, in our view, to decide on this point.
We shall describe a possible experiment to this end further down.

Note, in this respect, that a
Stern-Gerlach device is not the same as an optical polarizer. Even though
the two measurements are treated similarly in quantum mechanics, the polarization
measurement leaves the photon in some final state. A repetition of the
measurement can only confirm this state. The state of the particle or
an atom after an interaction with a magnetic field 
depends on the changes made by the measurement. 
In the current picture, these changes are reversibel: when 
the atom leaves the field, its properties are the same as before. 
In this case a repetition of the same deflection - which is what
an idealization in quantum mechanics predicts in every case - depends 
on an identical interaction process also for the second measurement. 
In case this is not observed, two explanations seem possible: (i) 
Decoherence effects between the atom and the experimental devices 
lead to a mixture of different spin states (this would be the explanation
in quantum mechanics), or (ii) the atomic electron is not polarized
at all (this is the theoretical prediction in the current model,
not including irreversible effects). It seems interesting to 
check these possibilities by high-precision measurements. 

Theoretically speaking, the change of an undefined combination of
particle states before the measurement to a defined particle state 
after the measurement is not a desirable feature of such an
experiment. Because, as already mentioned, it raises the problem of
the collapse of the wavefunction. Based on the treatment given in
this paper it seems that the problem could find its solution in the 
least expected way: there is, maybe, no collapse for this sort of 
measurement. All the more reason for a careful revision of 
Stern-Gerlach experiments.

The suggested theory can be checked experimentally by standard
Stern-Gerlach experiments. There are two distinct differences
between this model and the conventional model in quantum mechanics.

\begin{itemize}
\item
The first one concerns the forces of deflection in a single 
Stern-Gerlach measurement. In quantum mechanics this force 
depends only on the gradient of the field:

\begin{equation}
F_{QM} \propto \frac{\partial B_{z}}{\partial z}
\end{equation}

In the current model the force depends on the gradient of the field
{\it and} the amplitude of the field:

\begin{equation}
F_{MD} \propto B_{z} \frac{\partial B_{z}}{\partial z}
\end{equation}

The difference should be measurable in any high-precision 
measurement, since it concerns not so much absolute values 
but the dependency on the changes of magnetic fields.

\item
The second difference applies to multiple Stern-Gerlach experiments.
Provided the process really is reversible, the prediction differs 
significantly between quantum mechanics and this model. This effect
is far less significant, because decoherence will in practice destroy
the distinction between the states. But even then the difference should
be observable.
\end{itemize}

\section{Spin in quantum mechanics}\label{sec_spin}

In quantum mechanics the magnetic moment of a particle, and hence its spin,
is defined to comply with measurements. The measurements yield two possible
results, independent of any orientation of the magnet \cite{stern24}. The
experimental results therefore exclude a single valued function for the 
magnetic moment of a particle. Thus the classical picture, with its vector 
relation:

\begin{equation}
W = - {\bf \mu} {\bf B} \qquad {\bf \mu}, {\bf B} \in R^3
\end{equation}

is unsuitable to describe events at this level. We found in the preceding 
sections that even the two-valuedness could be a simplification. It takes only
two single points of the whole parameter space into account. This parameter
space is defined by the phase of the particle's intrinsic and the external
magnetic field. 

\subsection{Conservation laws and phase relations}

Spin cannot be interpreted as a rotation of an extended object around a
defined but hidden direction in space. Even though such a correspondence 
can be carried out to some extent (see for example Bohm and Hiley \cite{bohm93}),
it is ultimately insufficient. The reason is twofold: (i) The angular velocity
of a particle's surface exceeds the velocity of light. (ii) The degree of
freedom for the N-particle wavefunction of $2^{2N + 1}$ is much higher than
the physically relevant parameters of the system (proportional to $N$).

In our view, these aspects just point to different consequences of the 
same basic assumption: the motion of independent entities, with defined 
properties and existing independently of each other. From the viewpoint 
of quantum mechanics the total wavefunction
of a system is a Slater determinant of single particle wavefunctions.
This combination is commonly interpreted as a statistical feature of
quantum systems. We interpret it, slightly different, as the simplest
combination, which retains linearity throughout the system. It is 
therefore connected to the field properties of particles, and not
particularly to statistics. The interpretation bears on a shortcoming 
of conservation laws in classical mechanics. It
explains, why these entities (single particles) cannot be treated
independently from each other in a field theoretical context.
In this sense it explains, from a realistic and strictly physical
(as opposed to mathematical) point of view, why quantum mechanics
cannot be seen as a mechanical theory in any classical sense.

In general, conservation laws have two objectives: The first is to define
the standard of a formalization, or the zero point of a description. This
essentially Newtonian aspect (it is but a different form of his first
principle) considerably facilitates the description of a system, because
only changes of energy, momentum etc. need to be described. The second
objective is the identification of different objects with the same general
case. Every object with the same properties (energy, momentum etc.) can
be described with the same numbers, it is {\it the same object}. But
while this holds generally in classical mechanics, it is not true in
classical electrodynamics. Because in this case the phase of the field 
is generally relevant. So that even in the case when the field is finite,
the phase of intrinsic electromagnetic fields cannot remain unconsidered.
In a nutshell: an electron or photon with the same energy, linear momentum, and 
polarization as another electron or photon can not necessarily be described in
the same way. Not, if the phase along its path of propagation is
different. 
In this sense the identification in field physics is
more constrained than it is in mechanics. The additional constraint
is the phase relation. For this very reason the assumption in quantum
mechanics, that two objects \mbox{(= states)}, which differ only by a phase
$e^{i \varphi}$, are identical, is not generally valid. It makes a 
difference, if the phase of a single particle of a pair is changed by
$e^{i \varphi}$. Because in this case correlation measurements are
altered \cite{hofer00c}. Therefore the degeneracy with
respect to the phase of a particle's wavefunction is not a general
feature \cite{anastopoulos00}. And since this effect can be measured - and is
measured with high precision in quantum optics - the current 
framework of quantum mechanics, where the phase is seen as arbitrary
\cite{wigner59}, needs  to be modified. 

\subsection{Spin and interactions in quantum mechanics}

It is standard procedure to refer the mathematical representation of
spin via Pauli matrices to a necessary modification of the angular 
momentum \cite{greiner94}. The modification is due to the experimental
result of only two values for the magnetic interactions. It leads to
a modified Schr\"odinger equation, the Pauli equation:

\begin{equation}
\left[ \frac{1}{2 m} \left( \hat{\bf p} - e {\bf A}\right)^2 +
\mu_{B} \hat{\bf \sigma} {\bf B} \right] \Psi = 
i \hbar \frac{\partial}{\partial t} \Psi
\end{equation}

where $\mu_{B}$ is the Bohr magneton and $\Psi = (\psi_{1},\psi_{2})$ is
a two-spinor. The interaction in magnetic fields is now described by the 
term $\hat{\bf \sigma} {\bf B}$.

This description of spin via Pauli matrices is not limited to magnetic 
properties. It can be used quite generally to describe two-level systems.
In nuclear physics, for example, it is employed to describe the state
''proton'' and the state ''neutron'' of a nucleon. Therefore it appears
to be generally applicable. But in this case it cannot carry enough
physical significance, to make it analog to an angular momentum. And
then we may ask, whether it tells us anything more about a system than
precisely this: whatever you measure about spin, you always get exactly 
two symmetric results. The value of these results, for a single particle,
being defined to fit the measurements in atomic physics. And since all 
of these measurements are logically connected via the Pauli
(Schr\"odinger) equation, the definition is unique. The results we
have presented suggest that spin is only mathematically, but not
physically meaningful. It was proved that the interpretation of
spin as an angular momentum is wrong. And it was shown, how two
discrete solutions arise in every interaction of a particle
with a magnetic field. These discrete solutions are commonly
interpreted as ''spin states'' of fermions.

We are now in the position to understand, why spin, in quantum mechanics,
seems such a peculiar concept. In particular we can now try to answer
the questions, why it must be described by Pauli matrices; why these
spin-operators are linear and essentially abstract; and why they
must lead to two discrete solutions. Beginning with the first
question, we can say that: 

\begin{itemize}
\item
Since magnetic interactions affect the intrinsic features of a particle,
and since they lead, in a dynamic picture, to two symmetric solutions,
spin must be a non-unique and two-valued function. 
\end{itemize}

The rotational features of spin, most notably its $4 \pi$ symmetry,
also derive from the two-valuedness. The Pauli matrices
are a simple way to secure this property.

\begin{itemize}
\item
Because interactions are generally invariant to rotations of field
vectors in the system of a particle, spin cannot be described as
a two-valued vector in real-space.
\end{itemize}

The eigenvectors of spin must therefore be abstract (not real-space)
entities. Again, the spin matrices and the eigenspace related to it
just describe the simplest mathematical structures operating in such
a way. However, the physical process underneath these mathematical
entities is a completely different one. It is the continuous and dynamic
change of the field properties of particles, when they enter a magnetic
field, which is described in this abstract manner.
 
Also this problem, why quantum mechanics cannot describe
the physics behind its mathematical concepts, comes from the mechanical 
perspective. Because it is assumed in quantum mechanics that we can give an
integral description of the process by simply multiplying a
quality of the external field (${\bf B}$) with a quality of
the particle ($\hat{\bf \sigma} \cdot \psi$). Ultimately the problem 
comes from the assumed analogy with the magnetic moment. 
In microdynamics no such problem exists. The fields
remain perfectly defined during the whole interaction process. 
The dynamic picture thus gives a precise
spacetime account of interactions. A feature, which is impossible
to achieve in quantum mechanics. Not, because microphysics had 
any special properties to reckon with, but because quantum
mechanics uses unsuitable (mechanical) analogies.

\begin{itemize}
\item
The two symmetric results are the simplest degenerate cases for
magnetic interactions. They originate from the arbitrary phase of
the particles intrinsic and the external fields. In particular
from the existence of field components and kinetic components
of energy density.
\end{itemize}

In the general case the interaction is expected to exhibit a 
greater variety of possible results. How this general case is
to be formalized, and how interactions are to be calculated, 
remains to be described in future. The relevant message,
at this point, is only the following: 

Spin appears to be a simplification of particle properties. 
This simplification has its roots in the mechanical analogy.
Field theoretical models, considered by the author to be
generally superior, may change the way we describe microphysical
systems quite substantially. In addition, the whole model,
if verified, proves once more that quantum mechanics has achieved
only the fundamentals of the transition from macrophysics to
microphysics. In the long term it is thus more appropriate
to build upon a suitably modified field theoretical model. And
the theory of microdynamics seems to provide just such a model. 

\section{Conclusion}

In this paper we have given a precise spacetime account of magnetic
interactions. The model is based on the intrinsic properties of particles,
generally employed in microdynamics. We could show that homogeneous and
inhomogeneous magnetic fields in the simplest degenerate cases allow for
two symmetric solutions. This result is verified in experiments. We 
have analyzed the difficulties in quantum mechanics to describe these
measurements in a continuous model, and we found that they
arise from unsuitable mechanical analogies, in particular the
analogy of angular momentum. Finally, we pointed out possible 
experiments to check the model derived against the
standard model in quantum mechanics. 

\section*{Acknowledgements}

I am indebted to Guillaume Adenier and Robert Stadler for reading the draft
of the paper and providing me with valuable suggestions for its improvement.

%------------------------------------------------------------------------------%
%      R E F E R E N C E S
%
%------------------------------------------------------------------------------%

%------------------------------------------------------------------------------%
%      F I G U R E S
%
%------------------------------------------------------------------------------%

\end{document}